\documentclass[prb,twocolumn,aps,showpacs,showkeys]{revtex4}

\usepackage{graphicx}

\def\be{\begin{equation}}
\def\ee{\end{equation}}
\def\bea{\begin{eqnarray}}
\def\eea{\end{eqnarray}}

\begin{document}

\title{
Reverse Monte Carlo modeling of amorphous silicon 
}

\author{Parthapratim Biswas}
\email{biswas@phy.ohiou.edu}

\author{Raymond Atta-Fynn}
\email{attafynn@phy.ohiou.edu}

\author{D.~A.~Drabold}
\email{drabold@ohio.edu}

\affiliation{ 
Department of Physics and Astronomy, Ohio University, 
Athens, Ohio 45701, USA
}

\begin{abstract}

An implementation of the Reverse Monte Carlo algorithm 
is presented for the study of amorphous tetrahedral 
semiconductors. By taking into account a number of 
constraints that describe the tetrahedral bonding 
geometry along with the radial distribution function, 
we construct a model of amorphous silicon using the 
reverse monte carlo technique. Starting from a completely 
random configuration, we generate a model of amorphous silicon 
containing 500 atoms closely reproducing the experimental
static structure factor and bond angle distribution and
in improved agreement with electronic properties. Comparison
is made to existing Reverse Monte Carlo models, and the importance of
suitable constraints beside experimental data is stressed.

\end{abstract} 


\maketitle

\section{introduction}

The structure of amorphous semiconductors is well represented 
by continuous random network (CRN) model introduced by 
Zachariasen~\cite{zach} 70 years ago. The CRN model has the 
simplicity that each of the atoms should satisfy its local 
bonding requirements and should have as small strain as possible 
in the network which is generally characterized by having a narrow 
bond angle as well as bond length distribution. In spite of its 
apparent simplicity, the structural modeling of high quality 
tetrahedral amorphous semiconductors appears to be quite 
difficult. There have been many models of amorphous 
silicon~\cite{polk, gutt, woot,bark,Car,Thorpe,Stillinger,Kaxiras, Keating, Holender} 
proposed in the last 30 years which include from very simple 
hand-built model of Polk~\cite{polk}, computer generated periodic 
network model of Guttman~\cite{gutt} to the complex model of 
Wooten, Winer and Weaire (WWW)~\cite{woot}, but most of these 
models have some limitations in one way or another in describing 
the true nature of the amorphous state. The last method, the 
so called {\em sillium} approach of WWW, is based on the strategy 
of randomizing and relaxing the network is so far the most 
successful method of producing minimally strained CRN. The 
algorithm, in its modified form developed by Djordjevi$\acute{c}$ 
et.\,al.~\cite{Thorpe} and Barkema and Mousseau~\cite{bark}, can produce 
a CRN which is comparable to 
experimental results and is capable of producing a clean band 
gap without any defect states in the gap. 

In this paper we develop a different approach to model amorphous 
semiconductors known as reverse monte carlo (RMC) 
simulation~\cite{gere, laszlo, walt, rmc1, rmc2, rmc3, rmc4}. 
Our primary objective is to produce structural configurations 
that are consistent with experimental data but at the same 
time we go one step further to generate realistic configurations 
for comparison with models obtained via other routes.  We emphasize that
producing realistic models (meaning models which agree with all experiments) requires more than
spatial pair correlations, and identify additional constraints which lead to realistic
models.

The existing RMC models of amorphous 
semiconductors are found to be inadequate and 
fail to produce some of the basic experimental features of amorphous tetrahedral 
semiconductors.  Gereben \& 
Pusztai~\cite{gere, laszlo} have carried out RMC 
simulation of tetrahedral semiconductors using a number of models 
ranging from completely disordered configuration to randomized 
diamond structure. Although a certain degree of tetrahedral character 
in the bond angle distribution was reflected in their work, most 
of the models show an unphysical peak in bond angle distribution 
around $60^{\circ}$. The work of Walters and Newport~\cite{walt} on 
amorphous germanium made some progress toward getting the correct bond 
angle distribution, but the number of 3-fold coordinated atoms are 
quite high in their model and in absence of any discussion on 
local strain and electronic properties it is difficult to say how 
reliable their models are when it comes looking at the electronic 
properties. 

A developing area where RMC may be applied successfully is for
modeling amorphous materials exhibiting medium-range order (MRO). 
Such MRO is characterized by the existence of $10-20${\AA} 
scale structure. Recent developments in fluctuation 
electron microscopy (FEM)\cite{fem} and its application on 
amorphous germanium and silicon have indicated that computer 
generated CRN model of these materials lack the characteristic 
signature of MRO. Since RMC is based on experimental data, it 
provides a promising scheme to model amorphous materials having 
medium-range order by including the experimentally measured FEM 
signal as input data to augment pair correlations.

The plan of the paper is as follows. In section \ref{basic}, we 
briefly mention the basic philosophy of reverse monte carlo modeling 
and some of its salient features. This is followed by role of 
constraints in RMC modeling in section \ref{model} where we 
illustrate how a set of judiciously chosen constraints can be 
used to construct a reliable model of amorphous silicon. 
Finally we compare our results with those obtained from earlier 
RMC models and a model obtained via WWW algorithm. 

\section{Basics of RMC}
\label{basic}

The RMC method has been described in detail elsewhere~\cite{rmc4}. 
Here we briefly outline the basic philosophy of RMC. At the very 
basic level, RMC is a technique for generating structural 
configurations based on experimental data. The logic is very appealing: any model of a complex material
worthy of further study should, at a minimum, agree with what is known (that is, the
experiments). By construction, the RMC scheme enforces this (and for contrast, a 
molecular dynamics simulation may not). In an ideal
implementation, one should find a model agreeing with {\it all} known
information, but this is not easy to accomplish, though we make 
some progress below. The approach was originally 
developed by McGreevy \& Pusztai~\cite{rmc4, gere} for 
liquid and glassy materials for lack of different routes to 
explore experimental data but in recent years progress has 
been made toward modeling crystalline systems as well~\cite{keen}. 
Starting with a suitable configuration, atoms are displaced 
randomly using the periodic boundary condition until the input 
experimental data (either the structure factor or the radial 
distribution function) match with the data obtained from the 
generated configuration. This is achieved by minimizing a cost 
function which consists of either structure factor or radial 
distribution function along with some appropriately chosen 
constraints to restrict the search space. Consider a system 
having $N$ number of atoms with periodic boundary condition.
One can construct a generalized cost function for an arbitrary 
configuration by writing : 
\be
\label{eq-cost1} 
\xi=\sum_{j=1}^{K} \sum_{i=1}^{M} \eta_{i}^j \{F^j_E(Q_i)- F^j_c(Q_i)\}^2  
   + \sum_{l=1}^L \lambda_l P_l  
\ee
\noindent 

where $\eta_{i}^j$ is related to the uncertainty associated with 
the determination of experimental data points as well as the 
relative weight factor for each set of different experimental 
data. The quantity $Q$ is the appropriate generalized variable 
associated with experimental data $F(Q)$ and $P_l$ is the 
penalty function associated with each constraint.  For 
example, in case of radial distribution function and 
structure factor, $Q$ has the dimension of length and 
inverse length respectively. In order to avoid the atoms 
getting too close to each other, a certain cut-off 
distance is also imposed which is typically of the order 
of interatomic spacing. In RMC modeling, this is usually 
obtained from the radial distribution function by Fourier 
transform of the measured structure factor. This is equivalent 
to adding a hard sphere potential cut-off in the system 
which prevents the catastrophic build up of potential 
energy. 

In spite of the fact that RMC has been applied to many different 
types of systems -- liquid, glasses, polymer and magnetic 
materials, questions are often raised about the reliability 
of results obtained from RMC simulation. The method has never 
been accepted without some degree of controversy and the most 
popular criticism is the lack of unique solution from RMC. 
RMC can produce multiple configurations having 
the same pair correlation function. This lack of 
uniqueness, however, is not surprising, since
usually only the pair correlation function or structure factor is used in 
modeling the structure,  while there exists an infinite hierarchy 
of higher order correlation functions carrying independent structural information
are neglected. In other words, RMC samples from the space of 
all models consistent with some limited body of data -- in its simplest form
(analyzing a single experiment) RMC is an ideal gauge of how {\it non-specific}
the data is with respect to identification of an atomistic model. If the modeler
possesses {\it a priori} information independent of that implicit in the experiment 
being fit to, it is necessary to add this information to the modeling in some fashion to
receive a model in joint agreement with the experiment and the additional information.

\section{A new RMC model}
\label{model}

We begin by including the minimal information that is necessary 
to model a configuration of $a$-Si. In so doing, we use the radial 
distribution function obtained from a high quality model of 
amorphous silicon. This latter model was generated by Barkema 
and Mousseau~\cite{bark} using a modified form of WWW 
algorithm~\cite{woot} having bond angle distribution close to 
$10^{\circ}$ with 100\% 4-fold coordination. In addition to this 
RDF, we also impose the conditions that the average bond 
angle of all the triplets Si-Si-Si should be near $109.5^{\circ}$ 
and the corresponding root mean square deviation should be no 
less than $10^{\circ}$. The number of 4-fold coordinated atoms
is driven to a specified value during the simulation by including 
a constraint on the average coordination number. It is to be noted 
that while there is no limit to the number of constraints that 
can be included in the system, there is no guarantee that mere 
inclusion of more constraints will necessarily give better 
results. Forcing a completely random configuration with too many 
competing constraints may cause the configuration to be trapped in 
the local minimum of the function $\xi$ and may prevent the system 
from exploring a large part of the search space. By adding only 
the essential constraints that describe the chemical and geometrical 
nature of the bonding correctly, Eq.\,\ref{eq-cost1} can be written 
as : 

\bea
\xi & = & \sum_{i=1}^{M} \lambda_1\{F_E(x_i)- F_c(x_i)\}^2 \nonumber \\ 
    & + & \lambda_2\, (\theta_0 - \theta)^2  \nonumber \\ 
    & + & \lambda_3\, (\delta \theta_0 - \delta \theta)^2 \nonumber \\ 
    & + & \lambda_4 \{1 - \Theta(x - x_c )\} \nonumber \\ 
    & + & \lambda_5 \{\phi_0  - \phi\}^2
\label{eq-cost2}
\eea

where,

\bea
\theta &=& \frac{1}{N_{\theta}} \sum_{i\{j,k\}} \theta_{ijk} \nonumber \\
\delta \theta  &=& \sqrt{\langle {(\theta - \theta_0)^2}\rangle} \nonumber
\eea

In Eq.\,\ref{eq-cost2}, $\theta$ and $\delta \theta$ are 
the average angle and the rms deviation while $\phi$ 
and $\phi_0$ are the current and proposed concentration of 
the 4-fold coordinated atoms. It is important to note that each 
of the terms in Eq.\,\ref{eq-cost2} is non-negative, and should decrease ideally to zero 
during the course of minimization. 
\begin{figure}
\includegraphics[width=3.4in,height=3.4in,angle=270]{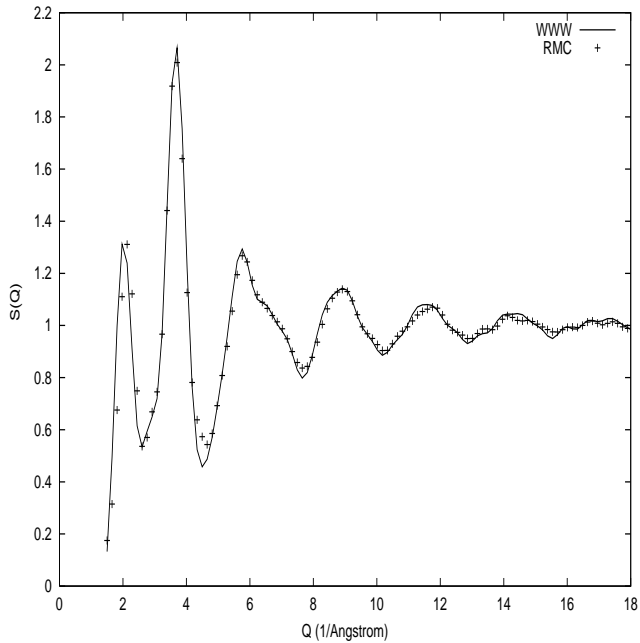}
\caption{ 
\label{cstr}
Structure factor obtained from a RMC (+) model containing 
500 atoms of $a$-Si. The solid line is obtained from a WWW 
sample of identical size and number density of atoms. 
}
\end{figure} 
Since the cost of energy associated with the bond length relaxation is 
more than the bond angle relaxation, atomic arrangements with large bond 
angle distribution but having correct RDF frequently result. The 
coefficients, $\lambda_1$ to $\lambda_3$, for the different terms in 
Eq.\,\ref{eq-cost2} can be chosen appropriately to minimize this effect. 
In general the coefficients $\lambda$ are constant during the course of 
simulation but the minimization procedure can be slightly accelerated 
by making them vary in such a way that the contribution from 
each of the term are of the same order during the course of simulation. 
The coefficient $\lambda_4$ is usually assigned a large value 
in order to include a hard sphere cut-off as mentioned earlier 
so that no two particles can come closer to $x_c$ while the 
coefficient $\lambda_5$ maintains the number of 4-fold 
coordinated atoms to a specified value. In RMC simulation of 
amorphous tetrahedral 
semiconductors one usually encounters the problem of having a 
pronounced peak at $60^{\circ}$. This peak is a characteristic 
feature of unconstrained RMC simulation and is due to the formation 
of equilateral triangles by three atoms. In the work of Gereben and 
Pusztai~\cite{gere,laszlo}, attempts were made to overcome this 
difficulty by constraining the bond angle distribution as well as 
by making an initial configuration which is 100\% 4-fold obtained 
from a diamond lattice. 
The resulting structure is, however, found to be unstable and on 
relaxation using a suitable potential, the configuration tends to 
get back toward the starting structure, i.e., randomized diamond 
in this case~\cite{note1}. 
\begin{figure}
\includegraphics[width=3.4in,height=3.4in,angle=270]{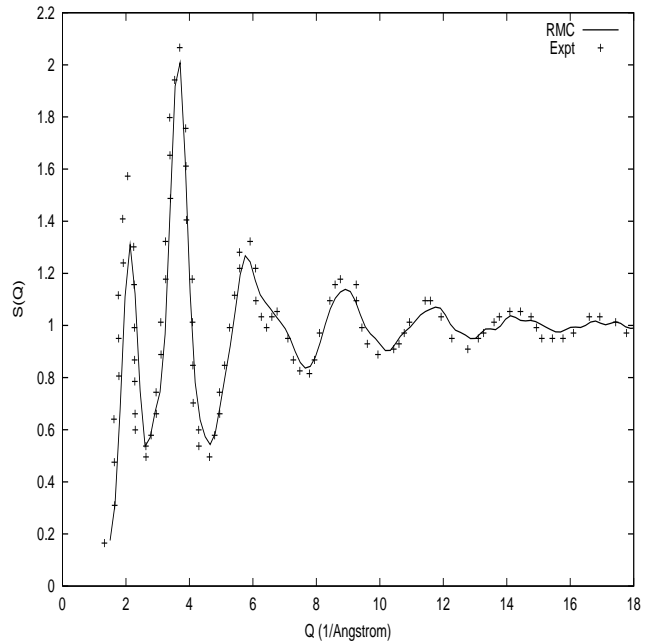}
\caption{ 
\label{str2}
Structure factor obtained for a 500-atom model of $a$-Si from 
RMC (solid line) and the experiment of Laaziri et al. as indicated 
in the figure. 
}
\end{figure} 
In the approach of Walter and 
Newport~\cite{walt}, the initial random configuration was examined 
and any ``triples'', i.e. , three atoms forming an equilateral 
triangle was removed before the beginning of RMC fit. By selective 
removal of such unwanted triplets, they have been able to generate 
configuration of $a$-Ge without having a peak at $60^{\circ}$. 
The approach that we have taken in our work is more general 
and starts with a completely random configuration. This 
eliminates, in the first place,  any possible local ordering 
that may exist in the starting structure (e.g., randomized 
diamond structure retains the memory of tetrahedral ordering). 
Furthermore, we have not included or excluded any special 
configuration in our starting structure, e.g., three atoms forming an 
equilateral triangle. Based on experimental consideration, we have 
included only the key features of 
amorphous tetrahedral semiconductors -- an average bond angle 
of $109.5^{\circ}$ having rms deviation of $10^{\circ}$ which 
is consistent with the RDF obtained from a WWW relaxed model 
used in our calculation. For the 500-atom model reported in 
this work, we have chosen a cubic box of length 21.18{\AA} 
which corresponds to number density 0.0526 atom/{{\AA}$^3$}. 
The initial configuration is generated 
randomly so that no two atoms can come closer to {2.0\AA}. The 
configuration is 
then relaxed by moving the atoms to minimize the cost function 
$\xi$. In addition to applying standard monte carlo moves in which a 
single or a group of atoms is randomly displaced, a variety of monte 
carlo moves have been implemented in our work. For example, in one 
of such moves, a 3-fold or 5-fold atom is selected and the 
nearest neighbor distance is examined. If the distance is greater 
than {2.7\AA}, the neighboring atom is displaced in order to bring 
the distance within a radius of {2.7\AA}. 
\begin{figure}
\includegraphics[width=3.4in, height=3.4in,angle=270]{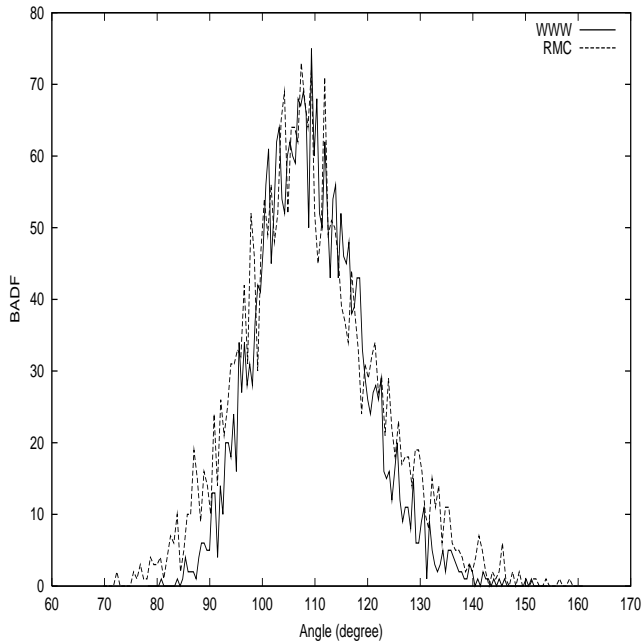}
\caption{
\label{cadf}
The bond angle distribution functions (BADF) for 500-atom $a$-Si 
from constrained RMC (dashed line) and WWW model (solid line). 
The rms deviation for the models are $12.5^{\circ}$ and $9.9^{\circ}$ 
respectively. 
}
\end{figure}
The maximum displacement of a monte carlo move is limited to 
0.2-0.4{\AA} throughout the simulation. Since we are interested 
in the electronic structure as well, we confine ourselves 
within a reasonable system size for studying the generated structure 
using a first principles density functional Hamiltonian. The density 
functional calculations were performed within the local density 
approximation (LDA) using the local basis first principles 
code {\sc Siesta}~\cite{siesta1}. We have used a non self-consistent 
version of density functional theory based on the linearization 
of the Kohn-Sham equation by Harris functional approximation~\cite{Harris} 
along with the parameterization of Perdew and Zunger~\cite{Perdew} 
for the exchange-correlation functional. 

\section{Results}
\begin{figure}
\includegraphics[width=3.35in, height=3.35in,angle=270]{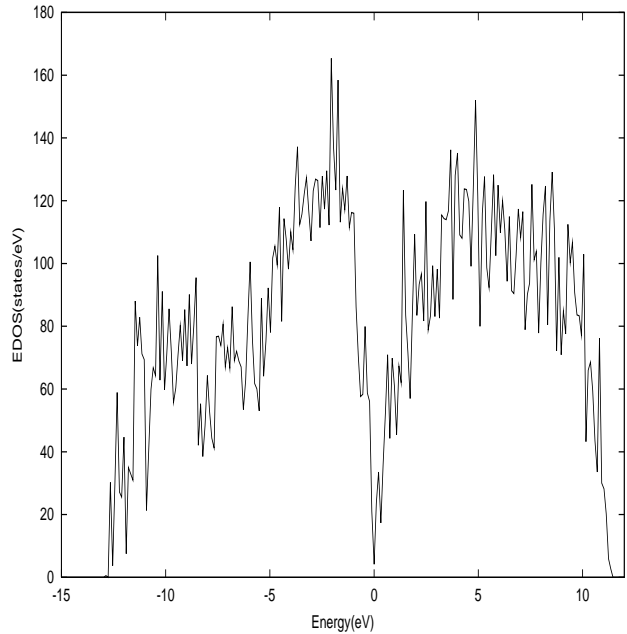}
\caption{
\label{edos}
The electronic density of states (EDOS) of 500-atom model of $a$-Si 
obtained from RMC simulation described in the text. The Fermi level is 
at E=0.
}
\end{figure}

The results for the model including all the constraints are presented 
in Figs.\,\ref{cstr}-\ref{edos}.
Since the structure factor is generally considered to be more sensitive to 
an arbitrary small change in the atomic positions than the radial distribution 
function, we have plotted the structure factors for the constrained RMC and 
WWW model in Fig.\,\ref{cstr}. It is evident from Fig.\,\ref{cstr} that the 
agreement between the RMC and WWW model is very good both for small and large 
values of Q. In order to further justify the credibility of our model, we have 
plotted in Fig.\,\ref{str2} the structure factor from the experiment of Laaziri 
et.\,al\cite{laaziri} along with the same obtained from our RMC model. 
Once again we find that the agreement between the structure factor 
from RMC and the experimental results is quite good except for the 
few points near the first peak. It is very tempting to think this 
deviation as a finite size effect coming from the finiteness of our 
model. We have therefore calculated the structure factor for WWW models 
containing 300 to 4096 atoms of Si but the deviation continues to remain. 
Holender and Morgan~\cite{Holender} also observed similar deviation near 
the first peak in their work with a much larger model containing 13824 
atoms which was compared with the experimental data obtained by Fortner 
and Lannin~\cite{Fortner}. 

In Fig.\,\ref{cadf}, we have plotted the bond angle distributions (BADF) 
for both the RMC and WWW model. As we have discussed in section II, the radial 
distribution function or structure factor can not alone provide all the 
necessary information that are needed to characterize an atomic configuration 
obtained from a reverse monte carlo simulation. A further characterization beyond 
pair correlation function is therefore vital and necessitates the need 
for getting some idea about the 3-body correlation function. It is clear from 
the Fig.\,\ref{cadf} that the distribution obtained from the RMC model follows 
the tetrahedral character observed in amorphous semiconductors. The average 
bond angle in this case is found to be $109.01^{\circ}$ with rms deviation of 
$12.5^{\circ}$. An important aspect of the bond angle distribution 
in Fig.\,\ref{cadf} is that most of the angles are lying between 
$70^{\circ}$-$150^{\circ}$ compared to $80^{\circ}$-$140^{\circ}$ in 
WWW case. We emphasize at this point that the earlier works on modeling 
amorphous tetrahedral semiconductors using RMC predicted a 
much wider bond angle distributions. Gereben and Pusztai~\cite{gere} 
have observed a pronounced, unphysical peak at $60^{\circ}$
except for the model starting with diamond structure while 
Walters and Newport~\cite{walt} have reported a bond angle 
distribution of $a$-Ge which is as wide as $60^{\circ}$-$180^{\circ}$. 
It is an important development here that by adding three more 
constraints ($\lambda_2, \lambda_3$, and $\lambda_4$) we have achieved 
a significantly improved results. 
Both the radial and the bond angle distribution functions 
reported here are at par with the results obtained from 
molecular dynamics simulation and is comparable 
to those obtained from WWW model. The fact that the inclusion 
of these two constraints leads to a significant improvement 
is not surprising. For a large continuous random network 
(CRN) model of amorphous tetrahedral semiconductor, one 
can approximate the bond angle distribution as nearly 
Gaussian~\cite{note3}. This approximated Gaussian distribution 
can defined by the first two moments of the distribution 
function. By specifying these two moments as constraints 
in Eq.\,\ref{eq-cost2}, we correctly describe the tetrahedral 
bonding geometry of the atoms which along with the radial 
distribution function produces a configuration more realistic 
than those obtained from models based on RDF or structure 
factor only. This suggests that in addition to the radial 
distribution of the atoms, one needs to include 
some relevant information about the nature of 3-body correlation 
among the atoms to construct a realistic configuration. 

Having studied the radial and bond angle distribution we 
now address the electronic density of states calculations. 
While the width of the bond angle distribution function (BADF) 
and the structure factor together indeed gives some idea about 
the quality of the model, some of the features e.g., the existence 
of spectral gaps and the position of defects states in the 
spectrum can be studied by looking at the electronic density of 
states only. The structure obtained from RMC simulation is first relaxed 
using the density functional code {\sc Siesta} and is 
found to be close to an energy minimum in the local density 
approximation (LDA). This is an important test for 
determining the stability of the structure obtained from 
RMC simulation and as far as we are concerned almost all 
earlier works on RMC have completely neglected this issue. 
In Fig.\,\ref{edos}, we have plotted the electronic density 
of states (EDOS) for the constrained model. The EDOS appears 
with all the characteristic features of $a$-Si with the 
exception of a clean gap in the spectrum. This behavior is not 
unexpected in view of the fact that 88\% of the total atoms are 
found to be 4-fold coordinated with an average coordination 
number 3.85. The presence of the defect states makes the gap 
noisy and at the same time the use of LDA underestimates the 
size of the gap. This EDOS is in significantly
better agreement with optical measurements than conventional
RMC models with much higher defect concentrations and
spurious bond angles. It is interesting to observe that the average 
coordination number from our model is very close to the experimental 
value of 3.88 reported by Laaziri et. al.~\cite{laaziri}.

\section{conclusions}

We have presented a model of amorphous silicon based on 
reverse monte carlo simulation. One of the novel features 
of our model is to start with a completely random structure 
and then to relax toward a realistic configuration by adding 
a number of physically relevant constraints. 
The characteristic features of the tetrahedral 
bonding are taken into account by adding constraints on 
average bond angle and its deviation from the mean while 
the number of 4-fold coordinated atoms is maintained at 
a specified value by further use of a constraint on average 
coordination number. The radial and the bond angle distribution 
obtained from our model is found to be in excellent agreement 
with a high quality CRN model produced by WWW algorithm. We 
have also compared the structure factor with the experimental 
data obtained by Laaziri et al. and observed a reasonably good 
agreement. By relaxing the model using the first principles density function 
code {\sc Siesta}, we find that the model is close to the 
energy minimum for LDA and is stable. The electronic density 
of states (EDOS) obtained from our model contains all the 
essential feature of amorphous silicon including a signature 
of the band gap. Although the model does not produce a clean gap 
in the spectrum, the quality of the EDOS is at par with 
models obtained from molecular dynamics simulation. Our RMC 
algorithm presents a significant improvement on previous RMC 
studies and makes it possible to compare for the first time, 
albeit qualitatively, the structural and electronic properties 
of RMC models with its WWW counterpart. We expect that further 
developments toward this direction will eventually make RMC as an 
useful modeling tool incorporating experimental information and 
can be used effectively without any criticisms in modeling 
complex materials. 

\acknowledgments
We thank John Abelson (UIUC) for motivating us in this work. We thank 
Gerard Barkema (ITP, Utrecht) for providing us with WWW models of 
amorphous silicon. We thank Stephen Elliott (Trinity College, 
Cambridge) for very valuable discussions. We acknowledge the support of 
National Science Foundation under grant Nos. DMR-0310933 and 
DMR-0205858.


\begin{thebibliography}{1}
\bibitem{note1} 
This observation is equally true in the context of original WWW model 
as pointed out by Winer and Wooten~\cite{ww} and was remedied in their 
work by making the number of bond switching per atom above a certain 
minimal value.
\end{thebibliography}
\end{document}